\documentclass[12pt]{iopart} 
\usepackage{graphicx}

\newcommand{\be}{\begin{equation}}
\newcommand{\ee}{\end{equation}}
\newcommand{\ba}{\begin{eqnarray}}
\newcommand{\ea}{\end{eqnarray}}
\newcommand{\Mc}{{\cal M}}
\newcommand{\Ms}{M_{\odot}}

\newcommand{\prd}{PRD }

\newcommand{\apj}{ApJ }

\newcommand{\bml}{\begin{mathletters}}
\newcommand{\eml}{\end{mathletters}}

\begin{document}

\title[A MCMC approach to the study of massive black hole binary systems with LISA]{A Markov Chain Monte Carlo approach to the study of massive black hole binary systems with LISA}

\author{E.D.L.Wickham, A.Stroeer, and A.Vecchio}

\address{School of Physics and Astronomy, 
University of Birmingham, Edgbaston, Birmingham B15 2TT, UK
}
\ead{edlw@star.sr.bham.ac.uk,astroeer@star.bham.ac.uk,av@star.sr.bham.ac.uk}
\begin{abstract}
The Laser Interferometer Space Antenna (LISA) will produce a data stream containing a vast number of overlapping sources: from strong signals generated by the coalescence of massive black hole binary systems to much weaker radiation form sub-stellar mass compact binaries and extreme-mass ratio inspirals. It has been argued that the observation of weak signals could be hampered by the presence of loud ones and that they first need to be removed to allow such observations. Here we consider a different approach in which sources are studied simultaneously within the framework of Bayesian inference. We investigate the simplified case in which the LISA data stream contains radiation from a massive black hole binary system superimposed over a (weaker) quasi-monochromatic waveform generated by a white dwarf binary. We derive the posterior probability density function of the model parameters using an automatic Reversible Jump Markov Chain Monte Carlo algorithm (RJMCMC). We show that the information about the sources and noise are retrieved at the expected level of accuracy without the need of removing the stronger signal. Our analysis suggests that this approach is worth pursuing further and should be considered for the actual analysis of the LISA data.
\end{abstract}

\section{Introduction}

The Laser Interferometer Space Antenna (LISA)~\cite{Bender98} is expected to detect gravitational waves from several massive black hole binary (MBHB) systems in the mass range $\sim 10^7\,\Ms - 10^4\,\Ms$ during its mission lifetime (see {\em e.g.}~\cite{Cutler2002a,Ses2005a} and references therein). Such sources produce a characteristic ``chirping'' waveform~\cite{Blanchet2002} that overlaps with the great variety of other signals in the LISA data stream, in particular quasi-monochromatic radiation generated by double white dwarf binaries (DWD)~\cite{HBW90,NYPZ04,MNS04} and the so-called extreme mass ratio inspirals~\cite{BC2004}. Radiation from MBHB's will be very strong and generate a signal-to-noise ratio in the range $\approx 10^2 - 10^4$~\cite{Cutler1998a,Hughes2002,Vecchio2004} that can easily dominate the detector output. It has been suggested that in order to extract full information on the other much weaker sources such strong signals need to first be removed using an ``identify and subtract'' analysis scheme. Some work was carried out to develop algorithms~\cite{CL03} to this end, although a full characterisation of their performance is not yet available. 

A quite different approach to LISA data analysis has recently emerged. It is primarily motivated by the large (and unknown) number of signals present in the data set and is formulated within the  framework of Bayesian inference. Such a scheme aims to construct {\em at once} the posterior probability density functions (PDF's) of the unknown model parameters given the data and the priors. A very effective numerical technique to construct the PDF's is based on the so-called Reversible Jump Markov Chain Monte Carlo algorithm (RJMCMC). The fact that the actual number of signals contained in the data set is unknown and needs to be determined yields a ``trans-dimensional'' problem, in which the number of search parameters is dynamically adjusted as a function of the signals considered in the model. Applications of (RJ)MCMC algorithms to (somewhat simplified) LISA data analysis problems -- either pure sinusoids or the monochromatic signals modulated by the LISA response -- have produced encouraging results\cite{Um2005a,Um2005b,Corn2005a}. 

In this paper we apply a MCMC algorithm to the study of overlapping etherogeneous sources. In particular, we concentrate on the issue of whether it is necessary to remove loud signals (such as those generated by MBHB's) to infer information about weaker sources by investigating whether one can {\em simultaneously} fit the data with a model containing both signals. In order to explore this issue, we consider a very simplified case in which LISA data contain two signals superimposed over the (instrumental) noise: one from a (strong) MBHB and one from a (weaker) DWD. Our approach is based on Bayesian inference implemented by means of an automatic Metropolis-Hastings RJMCMC sampler~\cite{Stroeer2006b} which tackles for the first time observations of in-spiral binaries in LISA. Not surprisingly, we find that this approach does return correct information on both signals; our results (although obtained by exploring a very limited parameter space) suggest that this approach is worth pursuing further and should be considered for the actual analysis of the LISA data. 

The paper is organised as follows: in Section~\ref{s:model} we introduce the simplified data analysis problem that we wish to tackle and the (RJ)MCMC approach used for the analysis; in Section~\ref{s:results} we discuss the results of the analysis and compare them to what is expected theoretically using the Fisher information matrix; Section~\ref{s:conclusions} contains our conclusions and pointers to future work.

\section{Bayesian approach to LISA observations}
\label{s:model}

In this section we discuss a simplified mock LISA data analysis problem to explore whether unknown signal parameters can be estimated simultaneously when strong sources are present by working in the framework of Bayesian inference. We also briefly introduce an automatic Metropolis-Hastings RJMCMC sampler~\cite{Stroeer2006b} that we have developed to tackle challenges in LISA data analysis. In our analysis we include some of the salient features of LISA data analysis, while keeping the problem at a simple level in order to ease the numerical implementation and limit the computational burden. 

We consider the output of a single Michelson observable (in the long wavelength approximation) that can be synthesised from the LISA constellation, following~\cite{Cutler1998a}. The mock data stream consists of the superposition of radiation from a MBHB and a DWD over a background noise. The gravitational waveform from the MBHB $h^{B}(t_j)$ is approximated at the leading Newtonian quadrupole order and can be cast in the form~\cite{Cutler1998a,Vecchio2004}:
\be
h^{B}(t_j) = A_0\, A_p(t_j)\,\left[\frac{f(t_j)}{f_L}\right]^{2/3}\, \cos\chi(t_j)\,,\quad\quad t_j \le t_c
\label{e:MBHB}
\ee
where $A_0$ is a constant amplitude, $A_p(t_j)$ the time dependent polarisation amplitude of the signal at the detector, $f(t_j)$ the instantaneous frequency of the signal at the (discrete) time $t_j$, $f_L$ a reference constant frequency -- that we set to the LISA nominal low-frequency cut-off $f_L = 10^{-4}$ Hz -- and $\chi(t_j)$ the phase at the detector. The signal is set to $h^B = 0$ for $t > t_j$. The MBHB signal depends on the vector of unknown parameters
\be
\vec{\theta}_{B}  = \left\{
A_0,\ \mathcal{M},\ t_c,\ \phi_0,\ \theta_N,\ \phi_N,\ \theta_L,\ \phi_L\right\}\,,
\label{e:vecMBHB}
\ee
where $\mathcal{M}$ is the chirp mass, $t_c$ the time at coalescence, $\phi_0$ an arbitrary constant reference phase and $\theta_N,\ \phi_N,\ \theta_L,\ \phi_L$ are the polar angles that describe the position of the source in the sky $\vec{N}$ and the orientation of the orbital angular momentum $\vec{L}$. We model the signal $h^{W}(t_j)$ from the DWD binary as monochromatic in the source frame; the LISA Michelson output can be written as~\cite{Cutler1998a,Vecc2004a}:
\be
h^{W}(t_j) =  A_0^{\prime}\, A_p^{\prime}(t_j)\, \cos\chi^{\prime}(t_j)\,;
\label{e:DWD}
\ee
the signal~(\ref{e:DWD}) depends on the vector of unknown parameters
\be
\vec{\theta}_{W} = \left\{
A_0^{\prime},\ f_0,\ \phi_0^{\prime},\ \theta_N^{\prime},\ \phi_N^{\prime},\ \theta_L^{\prime},\ \phi_L^{\prime}\right\} \nonumber
\label{e:vecDWD}
\ee
where $A_0^{\prime}$ and $f_0$ are the constant amplitude and emission frequency, respectively, $\chi^{\prime}(t_j)$ the phase at the detector, $\phi_0^{\prime}$ an arbitrary constant phase and $\theta_N^{\prime},\ \phi_N^{\prime},\ \theta_L^{\prime},\ \phi_L^{\prime}$ describe the vectors $\vec{N}^\prime$  and $\vec{L}^\prime$ for the DWD. The noise $n(t_j)$ is modelled in the time domain as a Gaussian and stationary random process with zero mean and unit variance, $\sigma^2 = 1$. The discrete datum $d(t_j)$ recorded by LISA can therefore be expressed as
\be
d(t_j) = h^{B}(t_j)+ h^{W}(t_j) + n(t_j)\quad\quad (j = 1,2,\dots,N)
\label{e:d}
\ee
where $N$ is the total number of points in the data set with sampling time $\Delta t$ and total observation time $T = N\,\Delta t$.

The goal of the analysis is to estimate the value of the unknown parameter vector (including noise) $\vec{\Theta}_k = \{\vec{\theta}_k,\sigma\}$ that describes the signal. Within the Bayesian framework, full information is contained in the joint posterior probability density function $p(k,\vec{\Theta}_k|\{d(t_j)\})$ of the (in principle unknown) model $m_k(t_j; \vec{\theta}_k)$, identified by $k$ and characterised by the parameter vector $\vec{\theta}_k$, given the data $\{d(t_j)\}$. Such a posterior PDF can be written using Bayes' theorem as
\be
p(k,\vec{\Theta}_k|\{d(t_j)\})=\frac{
p(k)p(\vec{\Theta}_k|k)p(\{d(t_j)\}|k,\vec{\Theta}_k)
}
{
\sum_{k'}\int p(k')p(\vec{\Theta}'_{k'}|k')p(\{d(t_j)\}|k',\vec{\Theta}'_{k'})\textnormal{d}\vec{\Theta}'_{k'}
},
\label{Bayestheorem}
\ee
where $p(k)$ and $p(\vec{\Theta}_k|k)$ are prior probabilities for the model labelled $k$ and its parameters $\vec{\theta}_k$, and $p(\{d(t_j)\}|k,\vec{\Theta}_k)$ is the likelihood of the data $\{d(t_j)\}$. The evaluation of the posterior~(\ref{Bayestheorem}) is notoriously difficult; in order to compute Eq.~(\ref{Bayestheorem}) we construct a single ``across-model'' Markov chain~\cite{Green2003a} with states $x=(k,\vec{\Theta}_k)$  to sample directly from the joint posterior $p(k,\vec{\Theta}_k|\{d(t_j)\})$, based on the knowledge of the likelihood and priors. For a given model the likelihood can be written as:
\be
 p(\{d(t_j)\}|k,\vec{\Theta}_k) \propto \sigma^{-N}\exp\left\{-\frac{1}{2\sigma^2}\sum_{j=1}^{N}\left[d(t_j) - m_k(t_j;\vec{\theta}_k)\right]^2 \right\}\,.
\ee
We adopt uniform prior distributions within given boundaries for all the parameters except noise, where we adopt priors proportional 1/$\sigma$.

In our Reversible Jump (RJ) implementation of the algorithm, the model $m_k(t_j; \vec{\theta}_k)$ is allowed to vary within a set of $k$ models. For the analysis reported in this paper, we set $k=1$ in the implementation and vary the model by hand in order to reduce the computational time.
The algorithm generates the (marginalised) posterior PDF's for each of the parameters of the chosen model and the noise, using chains with $10^8$ members which we have (empirically) tested is sufficiently long to ensure convergence to the target distribution. One of the main problems that we have encountered so far is the long turn-around time for the simulations: {\em e.g.} our code takes about 1.5 days to generate the posterior PDF's on the 16 parameters of the model that describe~(\ref{e:d}) using $N = 10^4$ data points on Tsunami, our 200 CPU (2GHz class processors) Beowulf cluster. A high priority for the immediate future is therefore to improve the efficiency of the algorithm.

\begin{table}
\begin{center}
\begin{tabular}{|l|ccccccccc|}
\hline
     & $A_0$ & $\theta_N$ & $ \phi_N$ & $\theta_L$ & $\phi_L$ &  $\phi_0$ & $\mathcal{M}$ & $t_c$ & $f_0$ \\
&&&&&&& $(\Ms)$ & (sec) & (Hz) \\
\hline
MBHB & 7.05 & 0.92 & 0.89 &  2.12 & 1.38 & 1.40    &  $1.9344\times 10^6$ &  $1.00001\times 10^6$ & -- \\
DWD  & 0.80  & 2.30 & 4.56 &  1.08 & 1.73 & 3.00    &         --           &  --   & $5\times 10^{-4}$  \\
\hline
\end{tabular}
\end{center}
\caption{Summary of the source parameters used in the simulations. The noise is modelled in the time domain as a stationary Gaussian process with zero mean and unit variance. We refer the reader to the text for more details, in particular Eqs.~(\ref{e:vecMBHB}) and~(\ref{e:vecDWD}) for the definition of the parameters.}
\label{t:param}
\end{table}

\section{Results}
\label{s:results}

\begin{figure}
\begin{center}
\mbox{
\scalebox{0.3}{\includegraphics{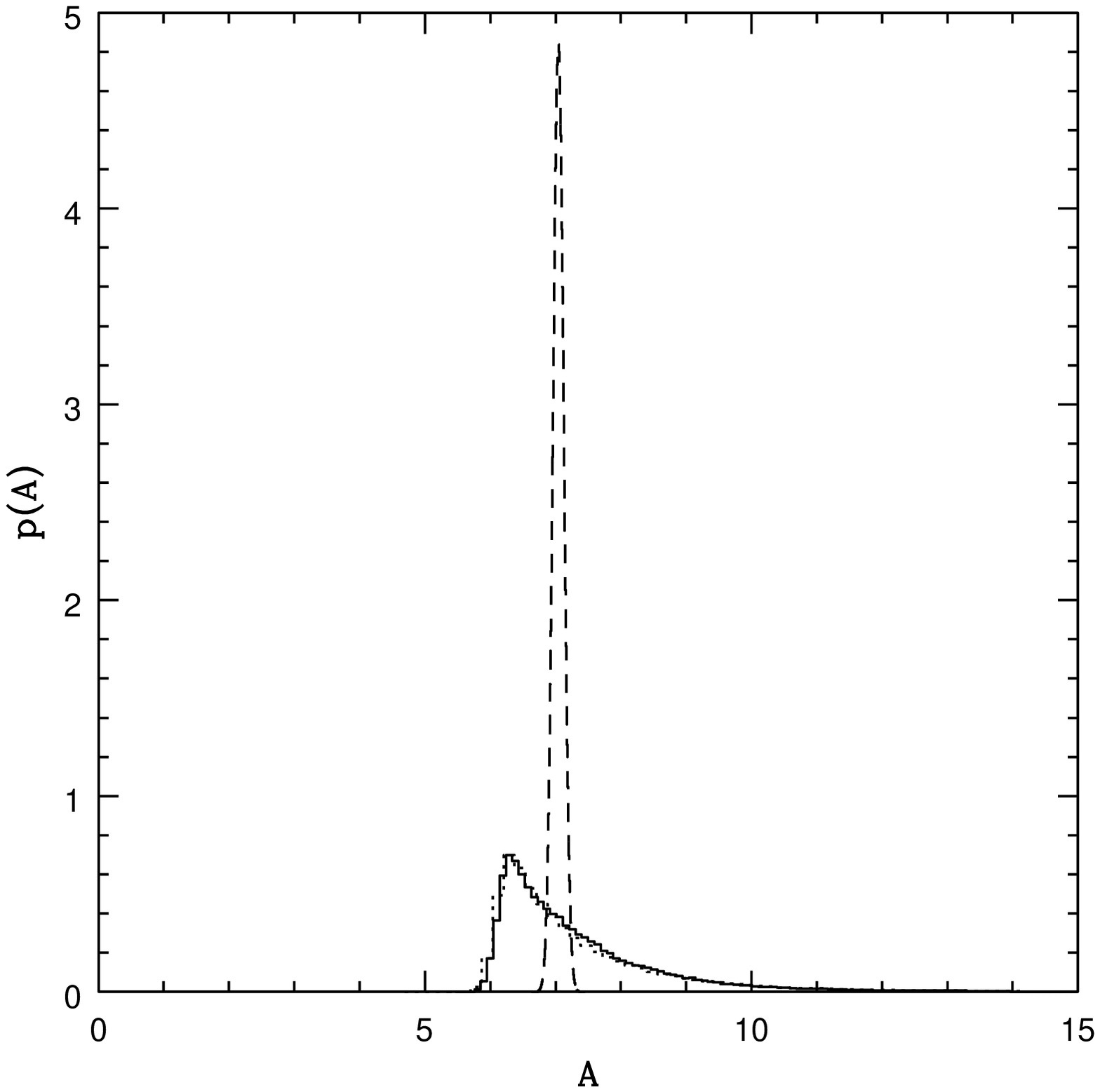}}
\scalebox{0.3}{\includegraphics{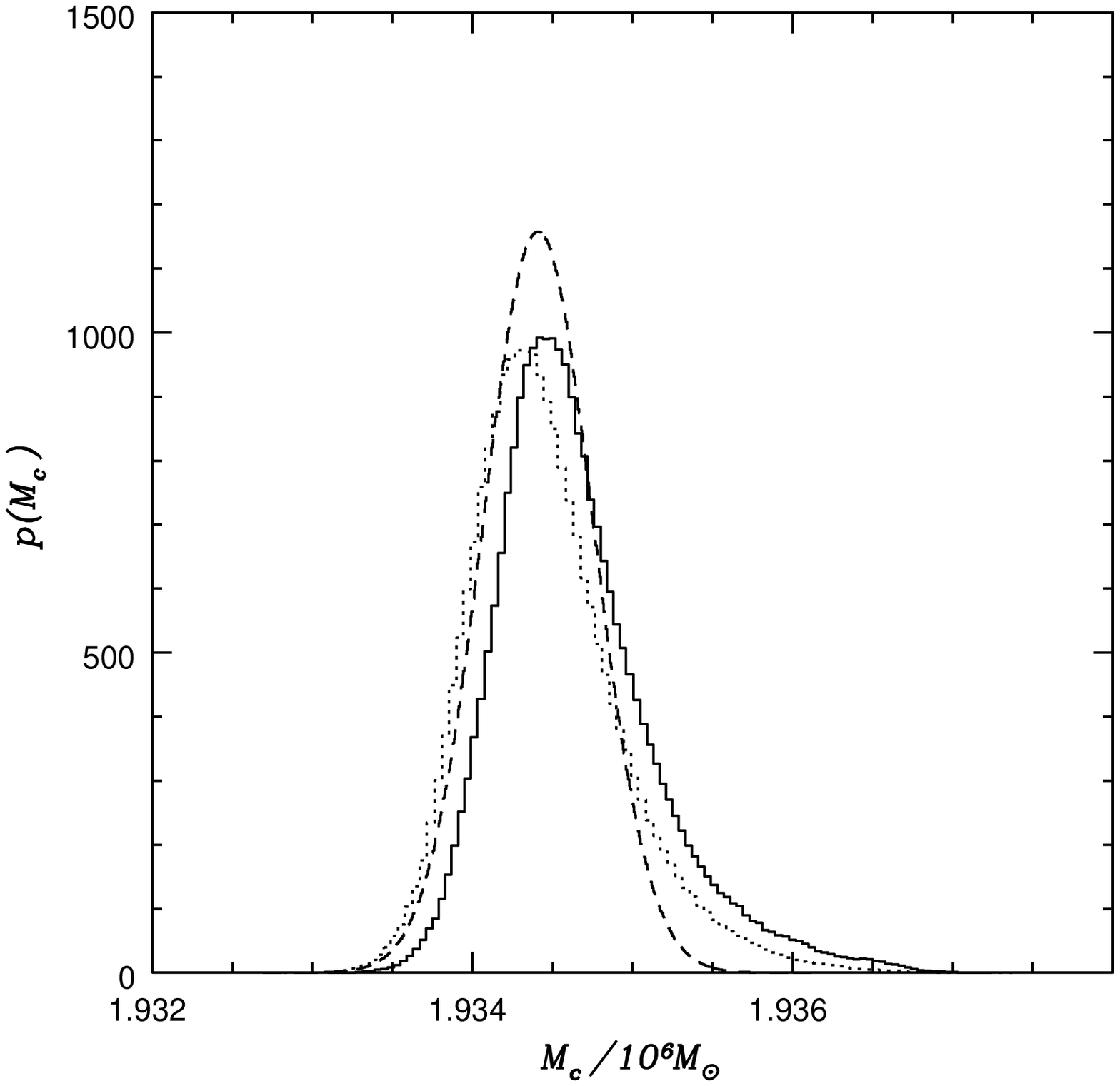}}
}
\mbox{
\scalebox{0.3}{\includegraphics{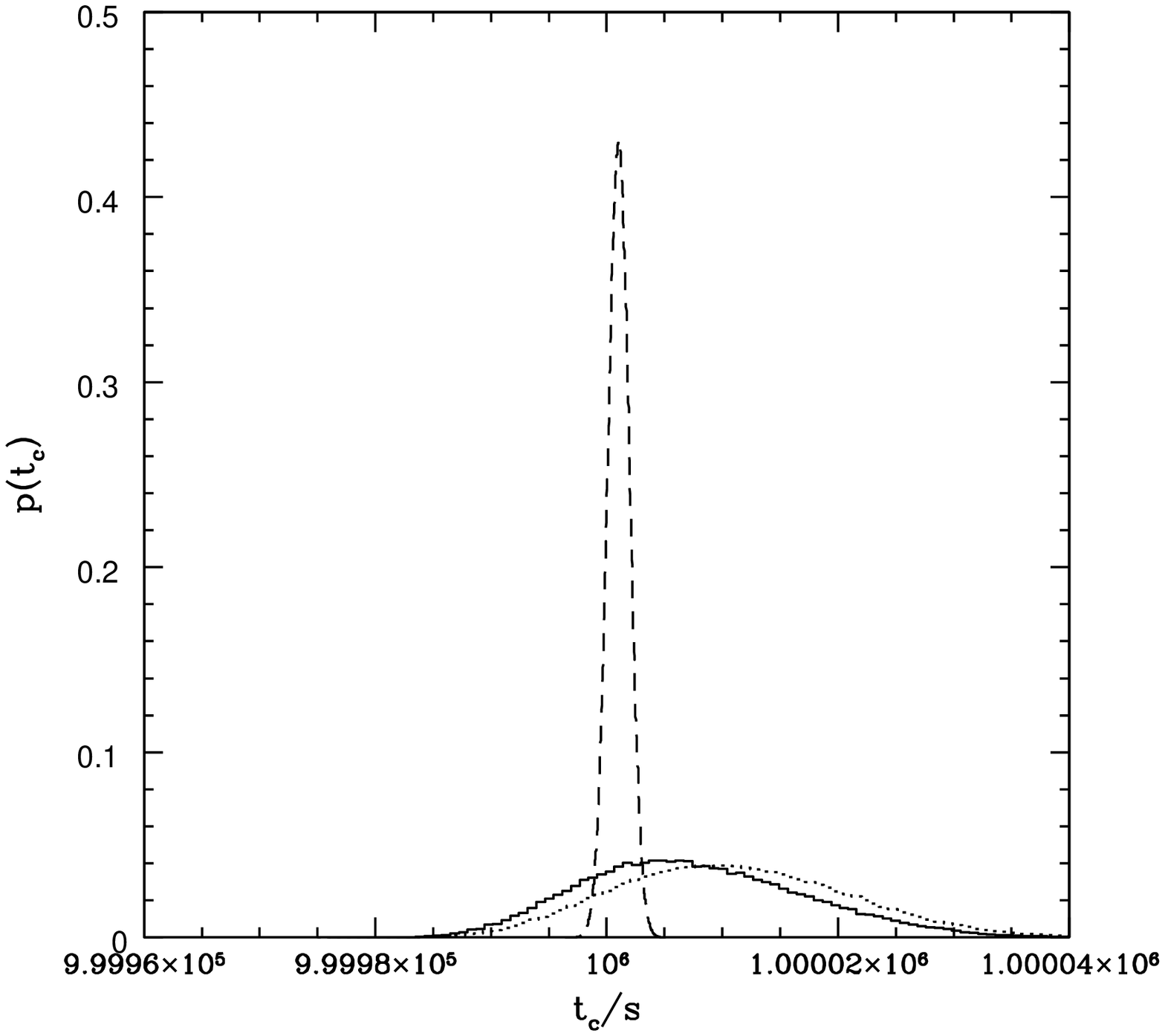}}
\scalebox{0.3}{\includegraphics{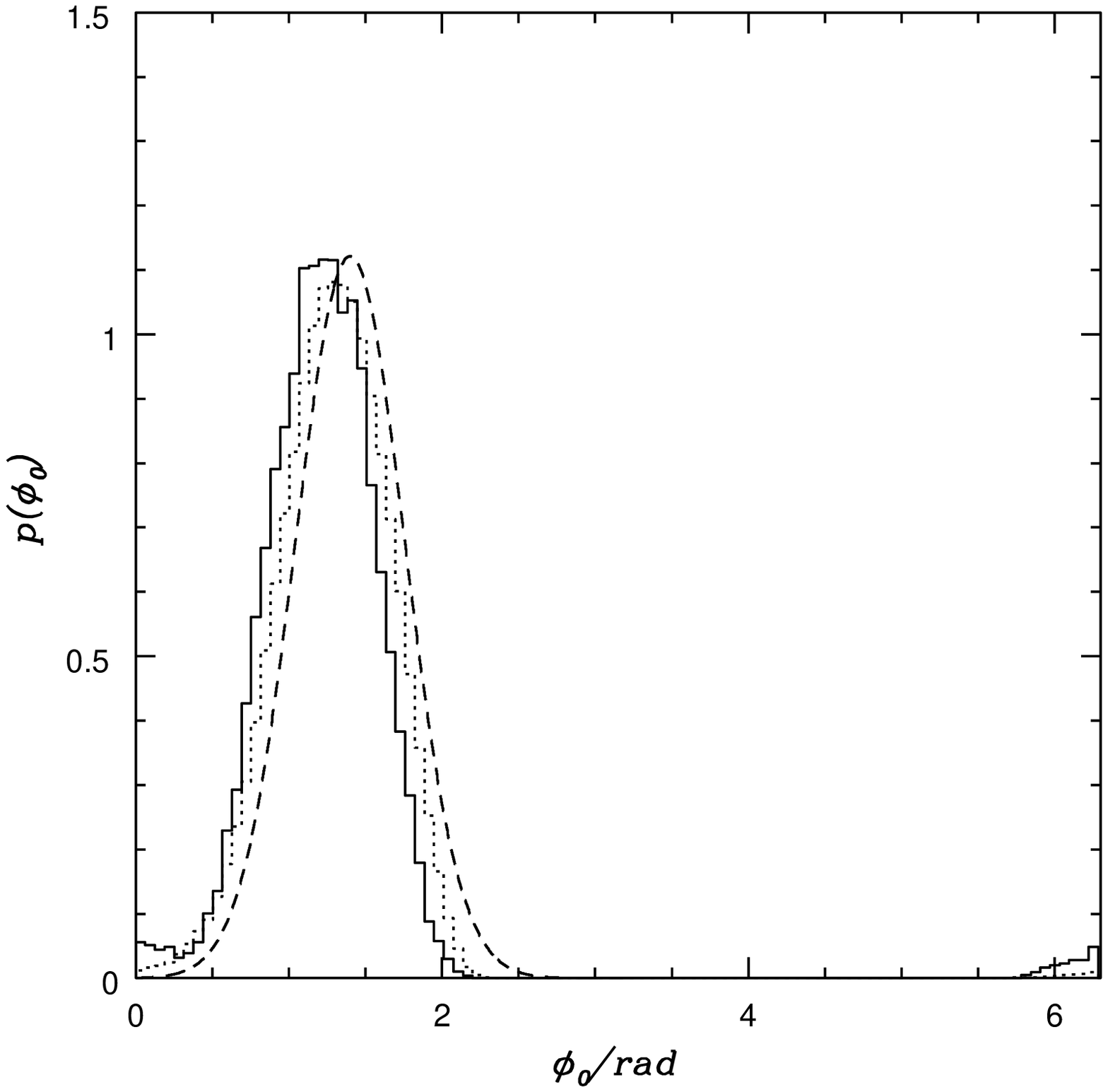}}
}
\caption{The posterior probability density functions for four of the parameters that describe the MBHB system:
amplitude $A_0$, chirp mass $\Mc$, coalescence time $t_c$ and initial phase $\phi_0$. The actual values of the signal parameters used in the simulation are given in Table~\ref{t:param}. The three lines correspond to the MCMC  results for the model $m_{B + W}$ (solid line), the model $m_B$ (dotted line), and the marginalised PDF's obtained from the multi-variate Gaussian distribution that one expects theoretically in the limit of signal-to-noise ratio $\rightarrow \infty$ (dashed line). In the latter we set the mean of the distribution to the exact value of the parameter chosen to inject a signal in the mock data set. See the text for more details.}
\label{Fig_BH}
\end{center}
\end{figure}

We generate in the time domain $T = 10^6$ seconds of LISA data with sampling time $\Delta t = 100$ sec, according to the model~(\ref{e:d}): the noisy data set contains an in-spiral signal from a MBHB just before coalescence and a quasi-monochromatic wave from a DWD. The source parameters are chosen as follows. We select the value of the chirp mass (set to ${\cal M} = 1.9344\times 10^6\,M_\odot$) in such a way that the (Newtonian) lifetime of the MBHB at frequency $f_L = 10^{-4}$ Hz coincides with the length of the observations. For numerical reasons, we actually set the instance of coalescence 1 s after the end of the data set. The signal spans the frequency range $10^{-4}$ Hz - 2.3 mHz. The frequency of the DWD is set to $f_0 = 5\times 10^{-4}$ Hz. The amplitudes of the signals are chosen so that the optimal signal-to-noise ratios correspond to $\rho_M\simeq 250$ and $\rho_D \simeq 45$ for the MBHB and the DWD, respectively. The location of the sources and orientation of the orbital angular momenta are chosen randomly. The exact values of the source parameters are listed in Table~\ref{t:param}. The two gravitational wave signals overlap completely in the time domain and (partially) in the frequency domain; note however that the overlap of the waveforms in parameter space (see {\em e.g.} Section 3 of~\cite{Vecc2004a} for a definition) is only $\sim0.009$. 

It is helpful to highlight the motivations of the simplifications that we have introduced in the simulation and the implications on the results. The choice of a shorter observation time with respect to the standard 1 yr of observation is determined by the computational costs: the time it takes to evaluate the PDF's scales linearly with $N$. The noise is chosen to be white for convenience of implementation. Note that the low frequency portion of the radiation for the MBHB becomes more important than it would actually be in LISA observations, because the noise does not rise below $\approx 1$ mHz. This assumption alters the actual numerical results of our analysis, but not the behaviour of the algorithm. We consider one Michelson output in the low frequency approximation (which is appropriate because the MBHB and DWD that we consider here radiate at frequencies below a few mHz) for simplicity and speed: the use of the full power of TDI (see {\em e.g.}~\cite{TD05} and references therein ) will affect the results from a quantitative, but not qualitative point of view. 

\begin{figure}
\begin{center}
\mbox{
\scalebox{0.3}{\includegraphics{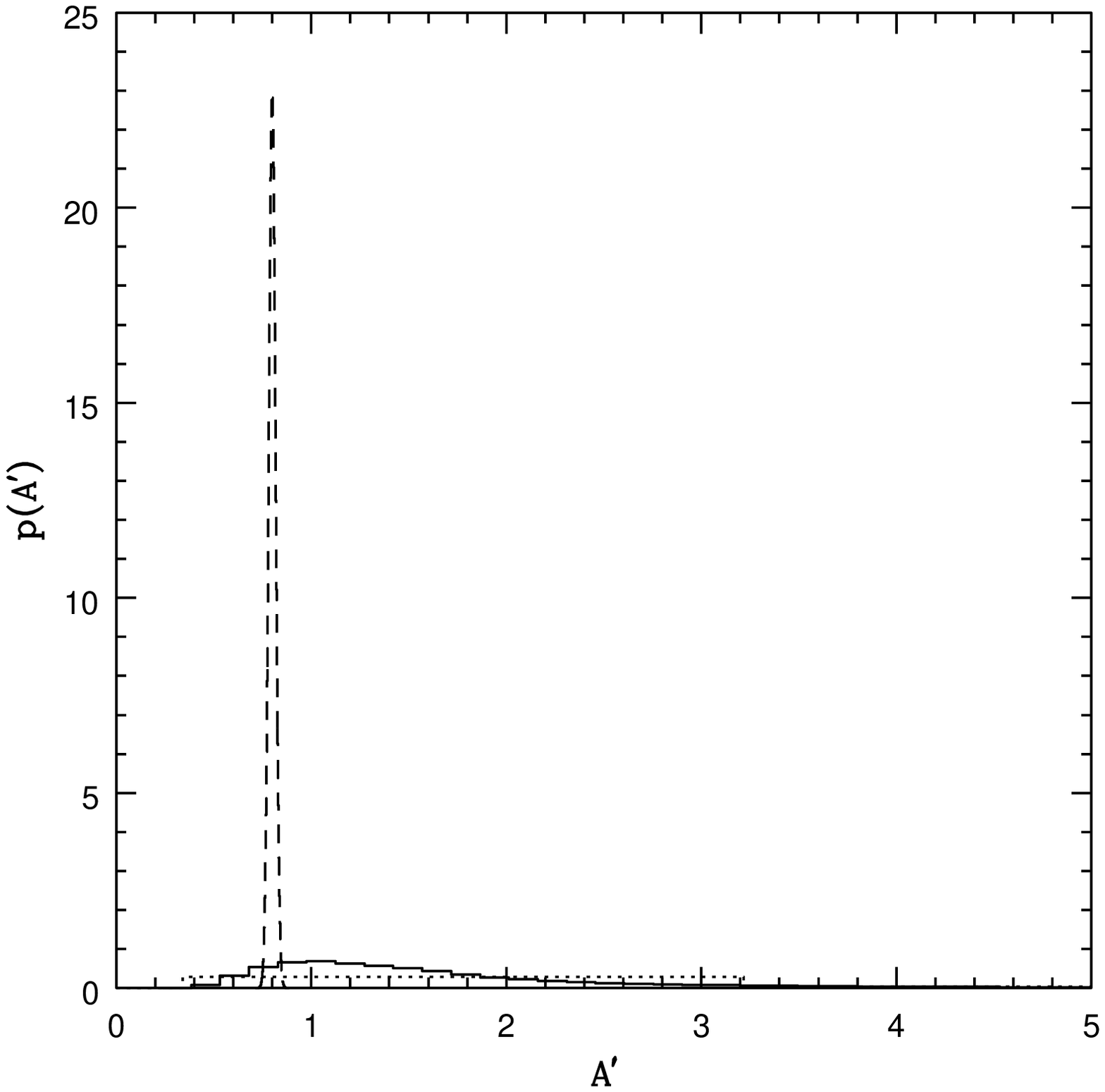}}
\scalebox{0.3}{\includegraphics{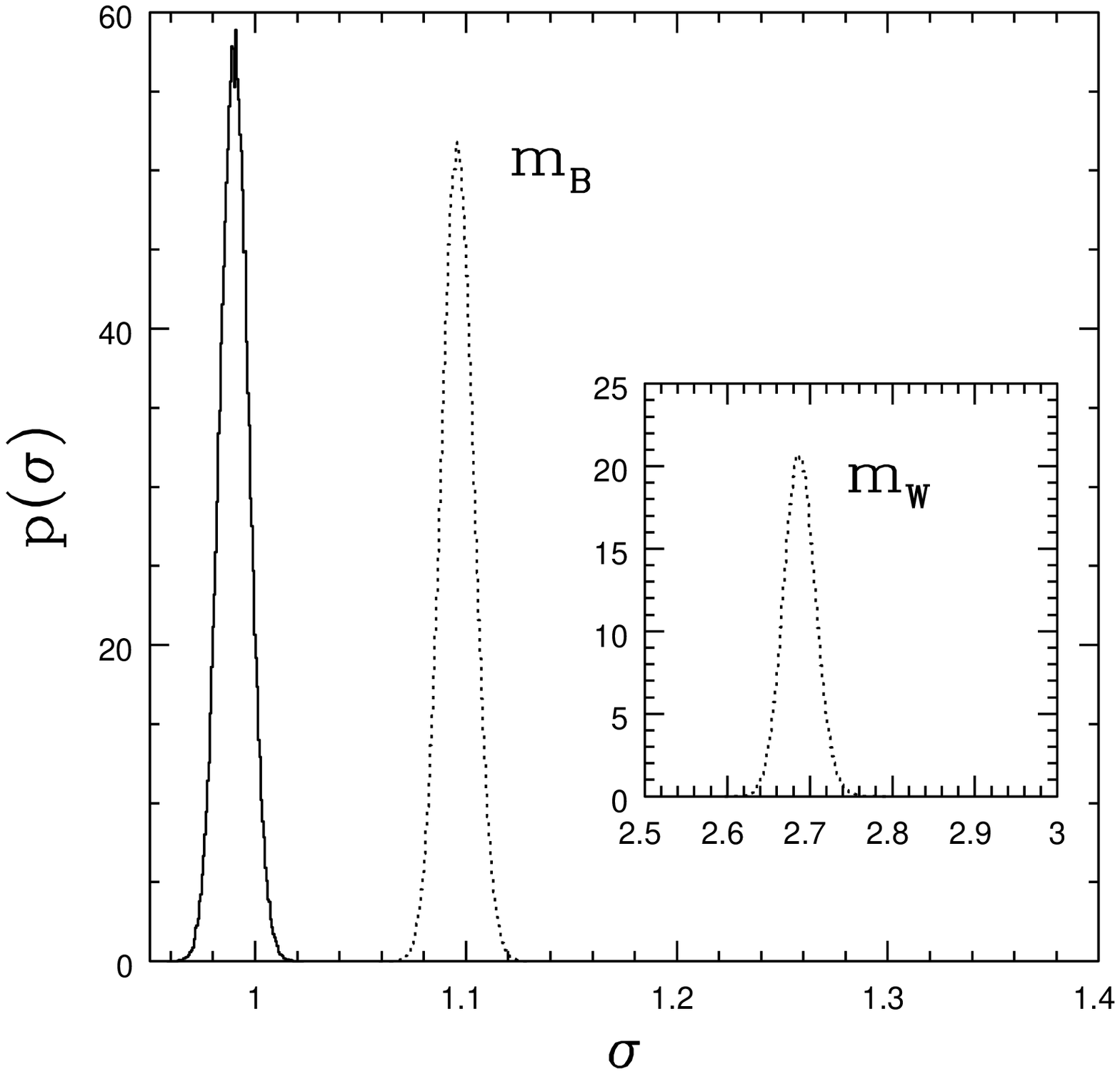}}
}
\mbox{
\scalebox{0.3}{\includegraphics{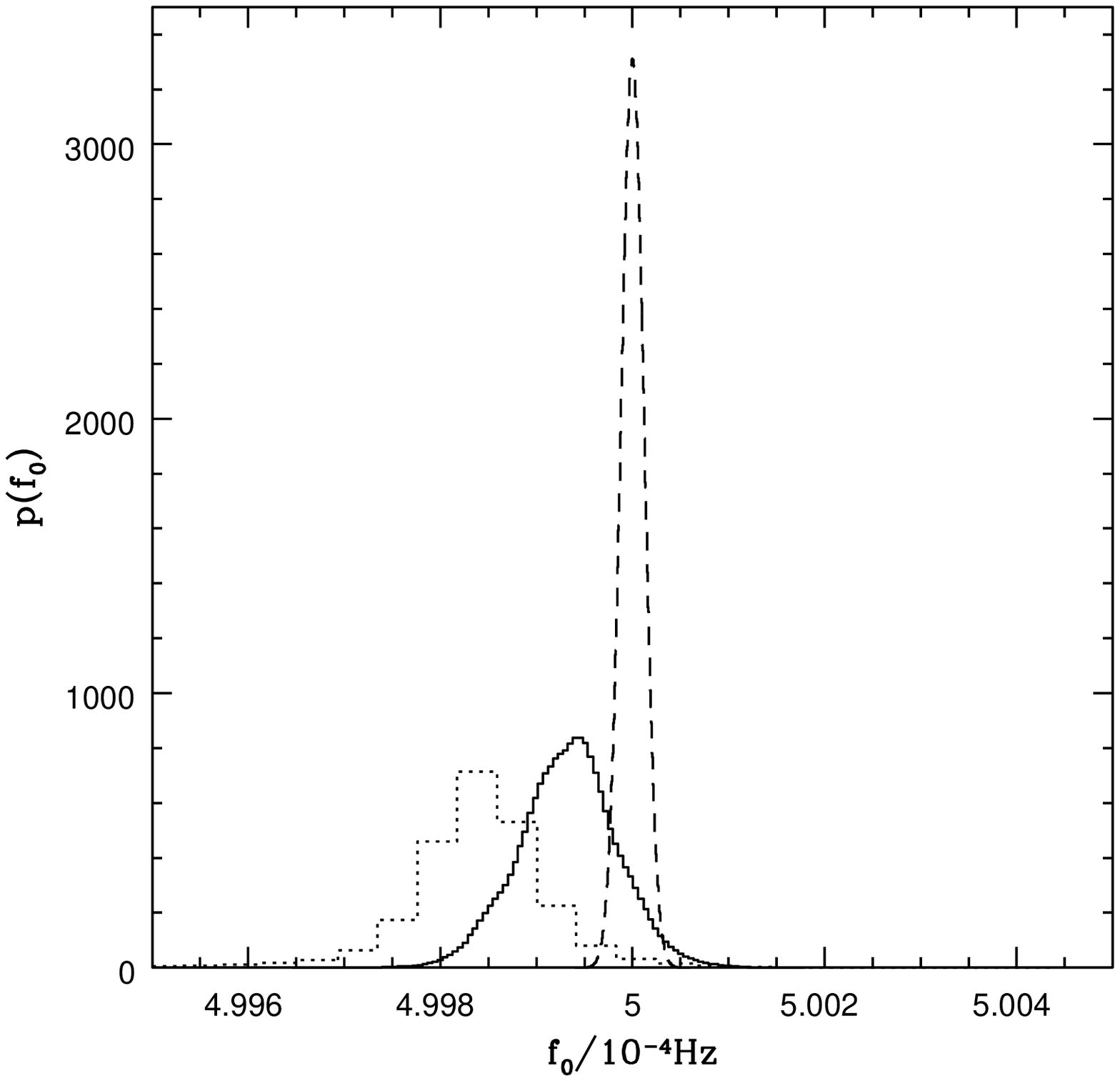}}
\scalebox{0.3}{\includegraphics{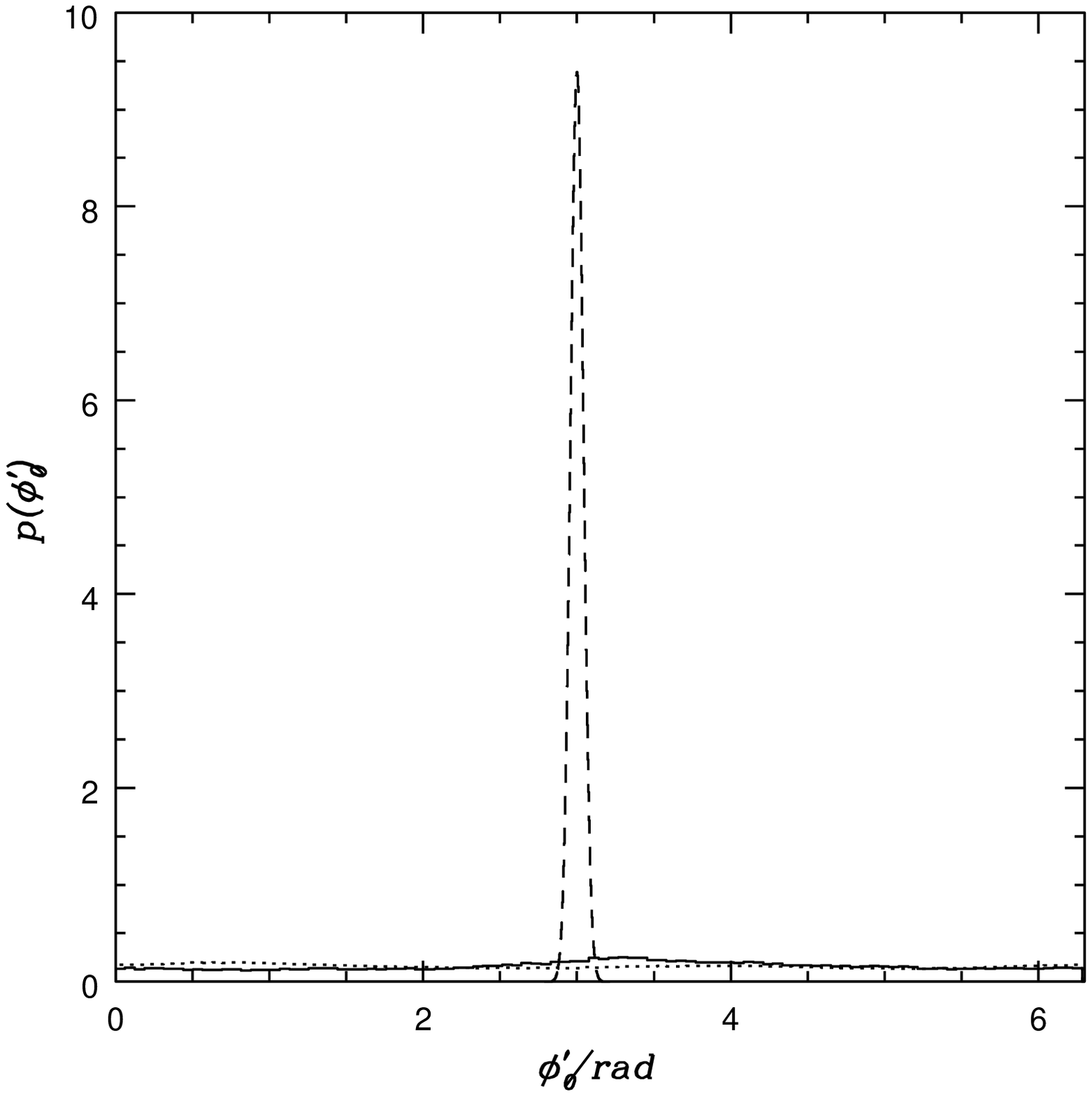}}
}
\caption{The posterior probability density functions for three of the DWD parameters -- amplitude $A_0'$, frequency $f_0$ and initial phase $\phi_0'$ -- and the variance of the noise $\sigma$. The actual values of the DWD parameters used in the signal injection are reported in Table~\ref{t:param}. The three lines in the plots for $A_0'$, $f_0$ and $\phi_0'$ correspond to the MCMC results for the model $m_{B + W}$ (solid line), the model $m_W$ (dotted line), and the marginalised PDF's obtained from the multi-variate Gaussian distribution that one expects theoretically in the limit of signal-to-noise ratio $\rightarrow \infty$ (dashed line). In the latter we set the mean of the distribution to the exact value of the parameter chosen to inject a signal in the mock data set. For the PDF of $\sigma$ the solid line corresponds to the model $m_{B + W}$ and the dotted lines to the case $m_W$ and $m_B$. In the computation of the theoretical multivariate Gaussian distribution of the source parameters the noise level is assumed to be known.}
\label{Fig_WD+noise}
\end{center}
\end{figure}

\begin{figure}
\begin{center}
\mbox{
\scalebox{0.3}{\includegraphics{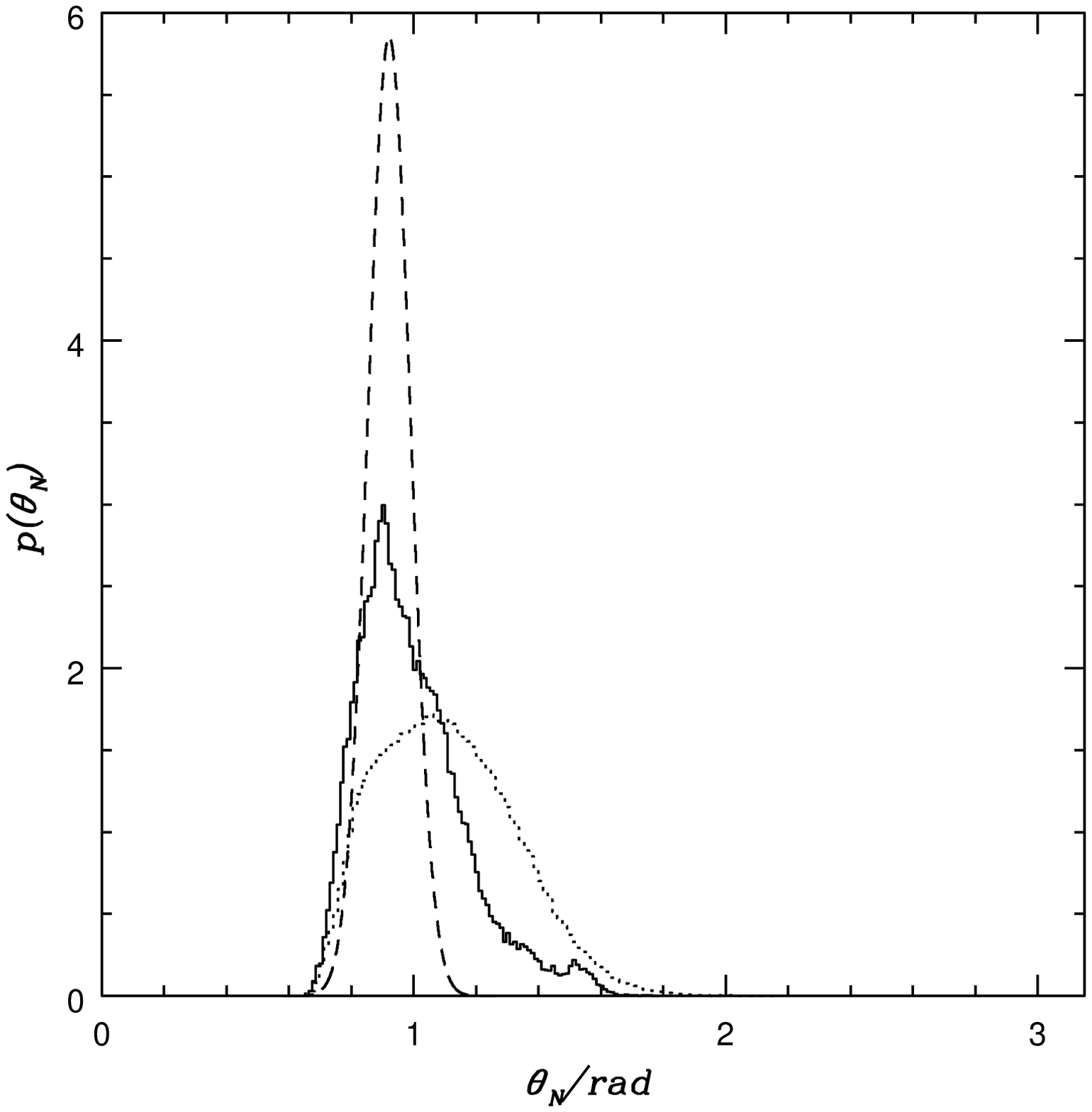}}
\scalebox{0.3}{\includegraphics{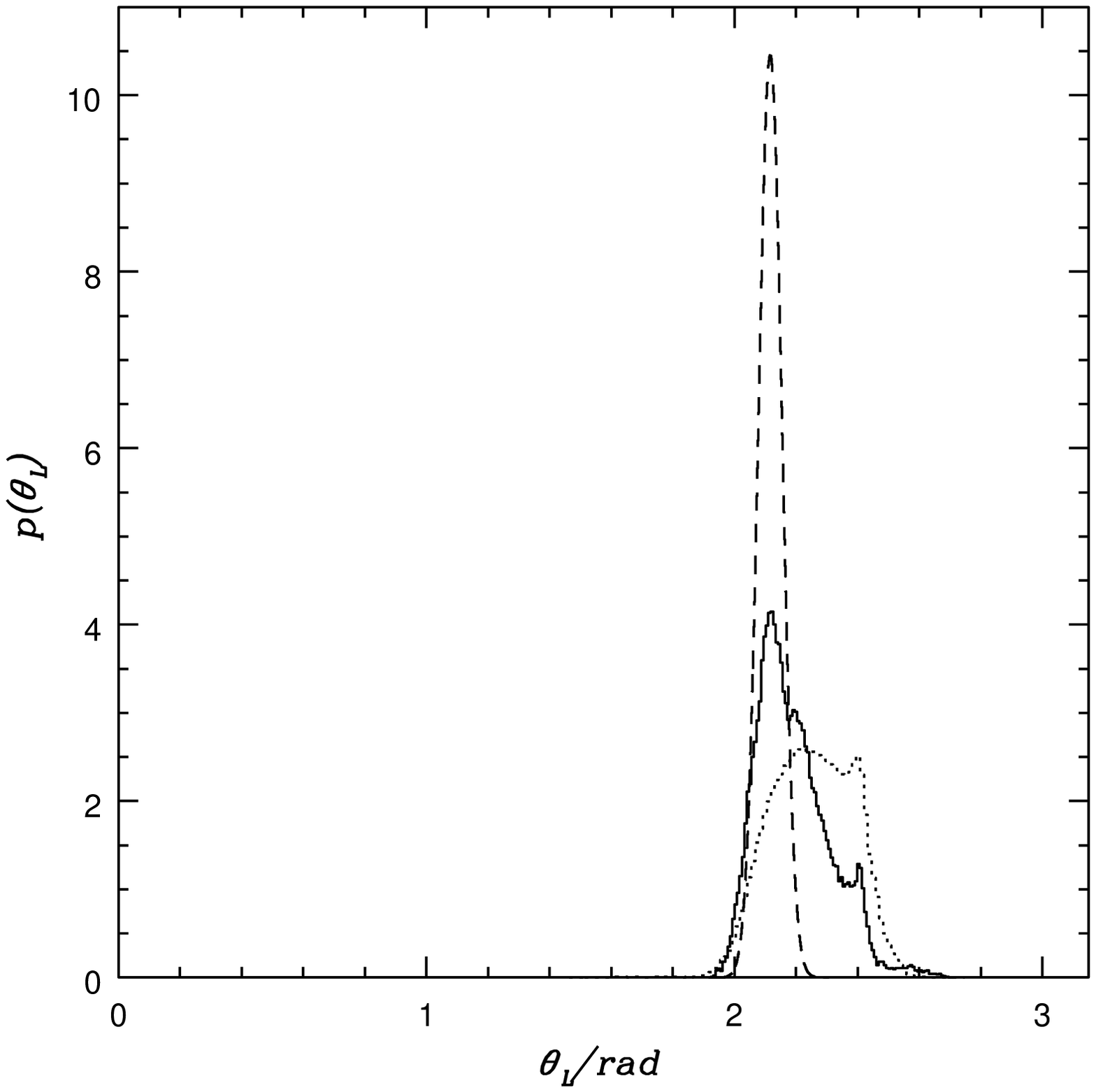}}
}
\mbox{
\scalebox{0.3}{\includegraphics{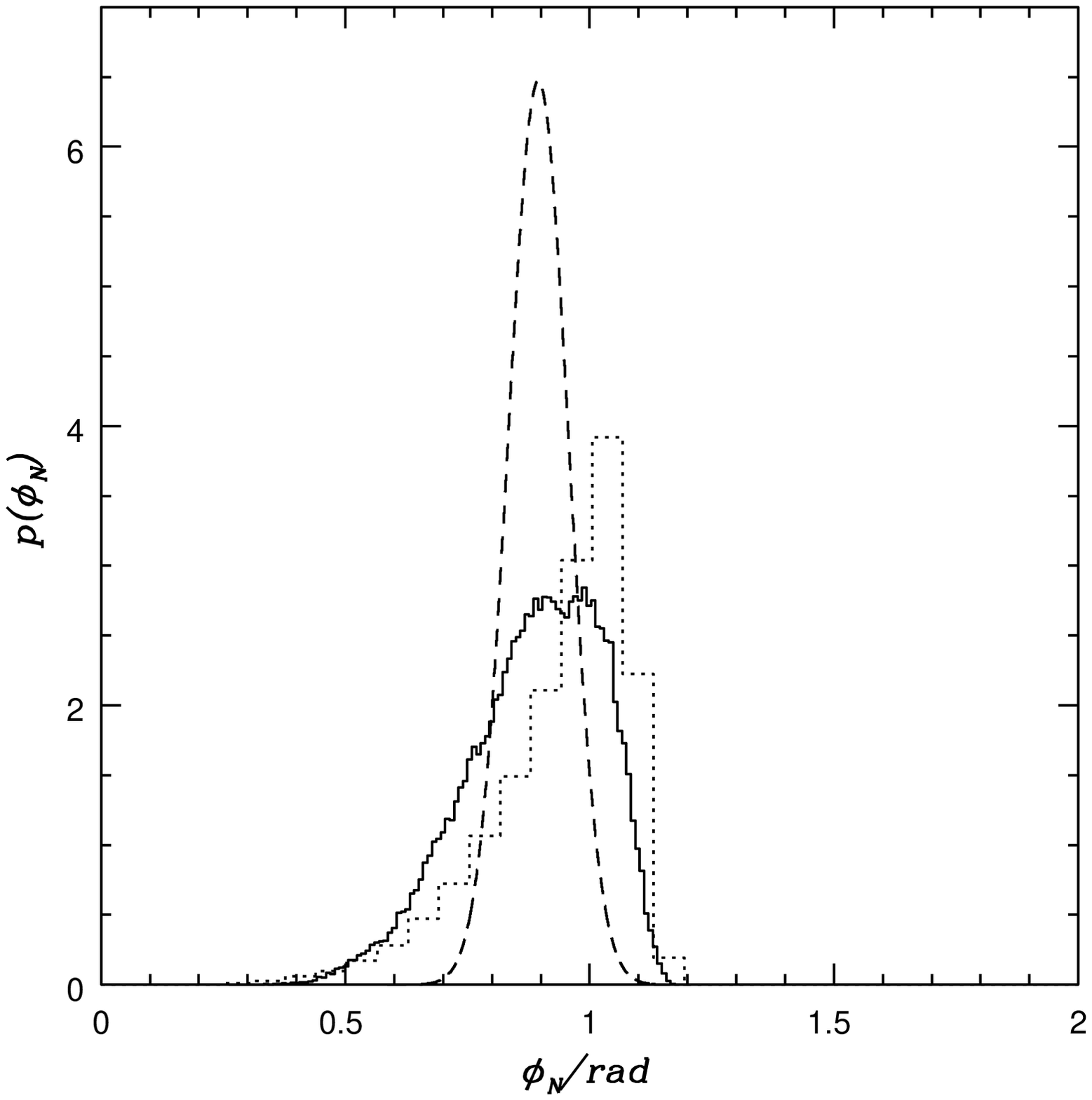}}
\scalebox{0.3}{\includegraphics{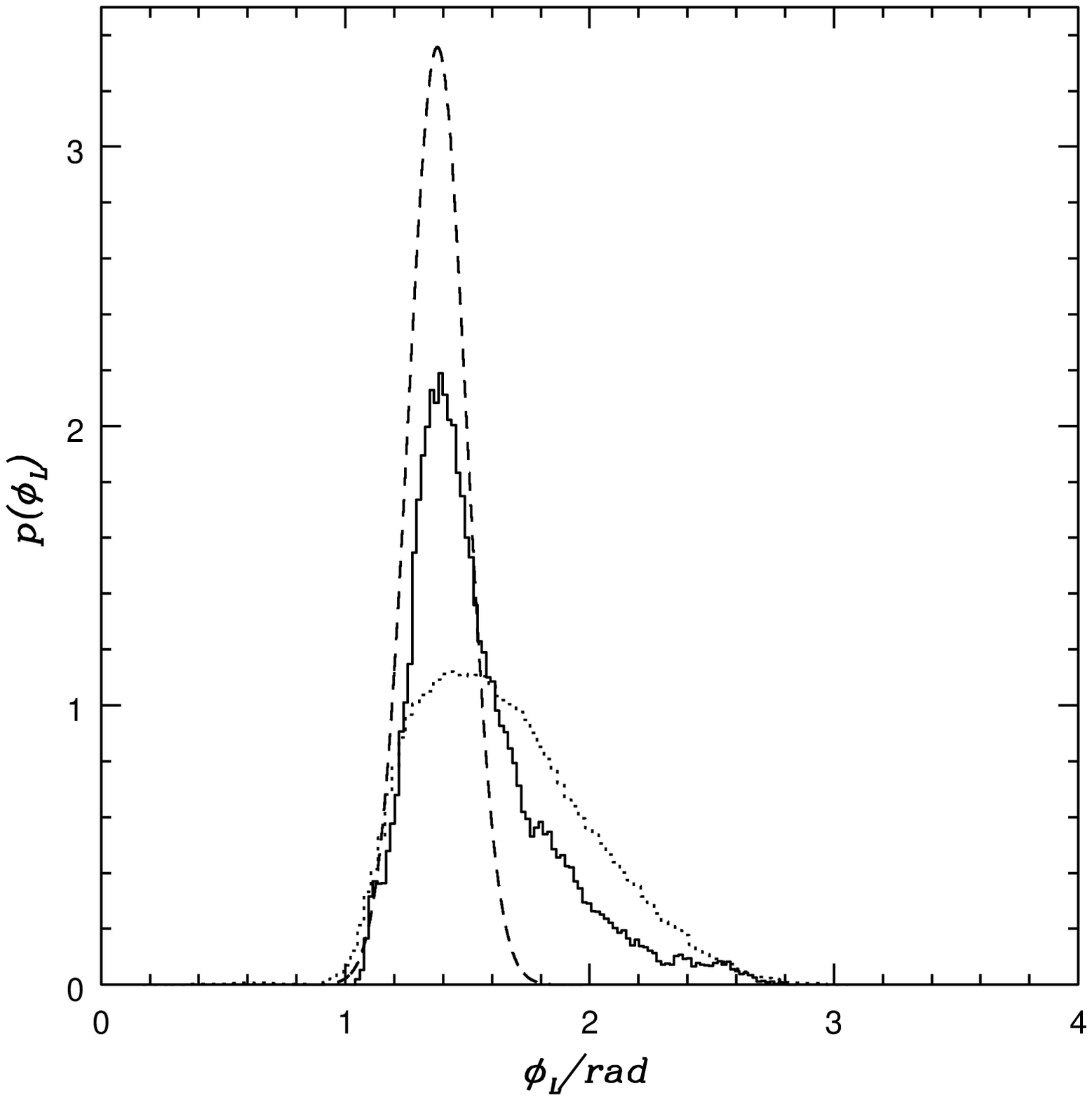}}
}
\caption{Same as for~\Fref{Fig_BH} but for the angular parameters describing the position $\hat N$ and orientation $\hat L$ of the MBHB. The actual parameter values are given in Table~\ref{t:param}.}
\label{Fig_BHangles}
\end{center}
\end{figure}

\begin{figure}
\begin{center}
\mbox{
\scalebox{0.3}{\includegraphics{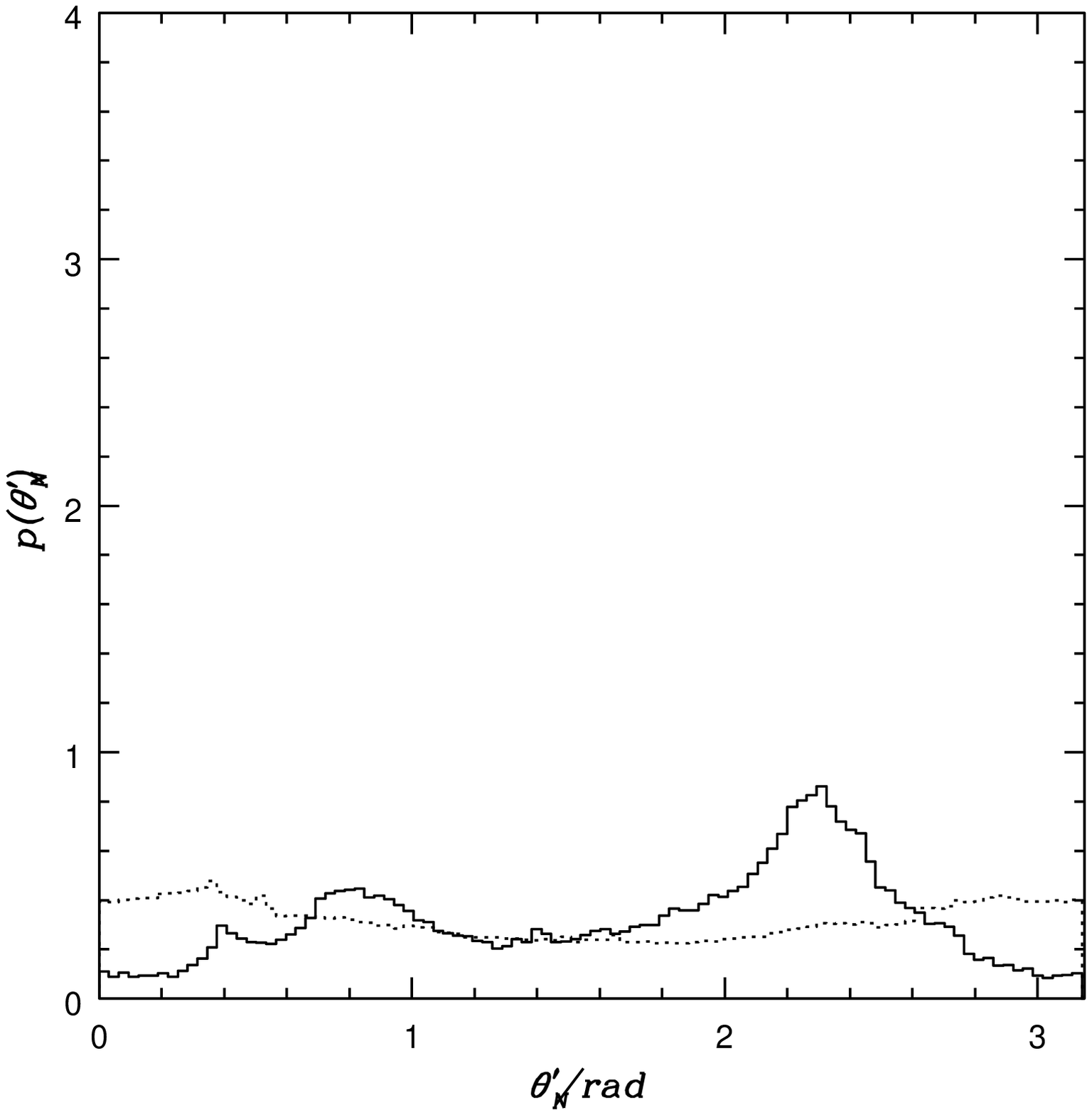}}
\scalebox{0.3}{\includegraphics{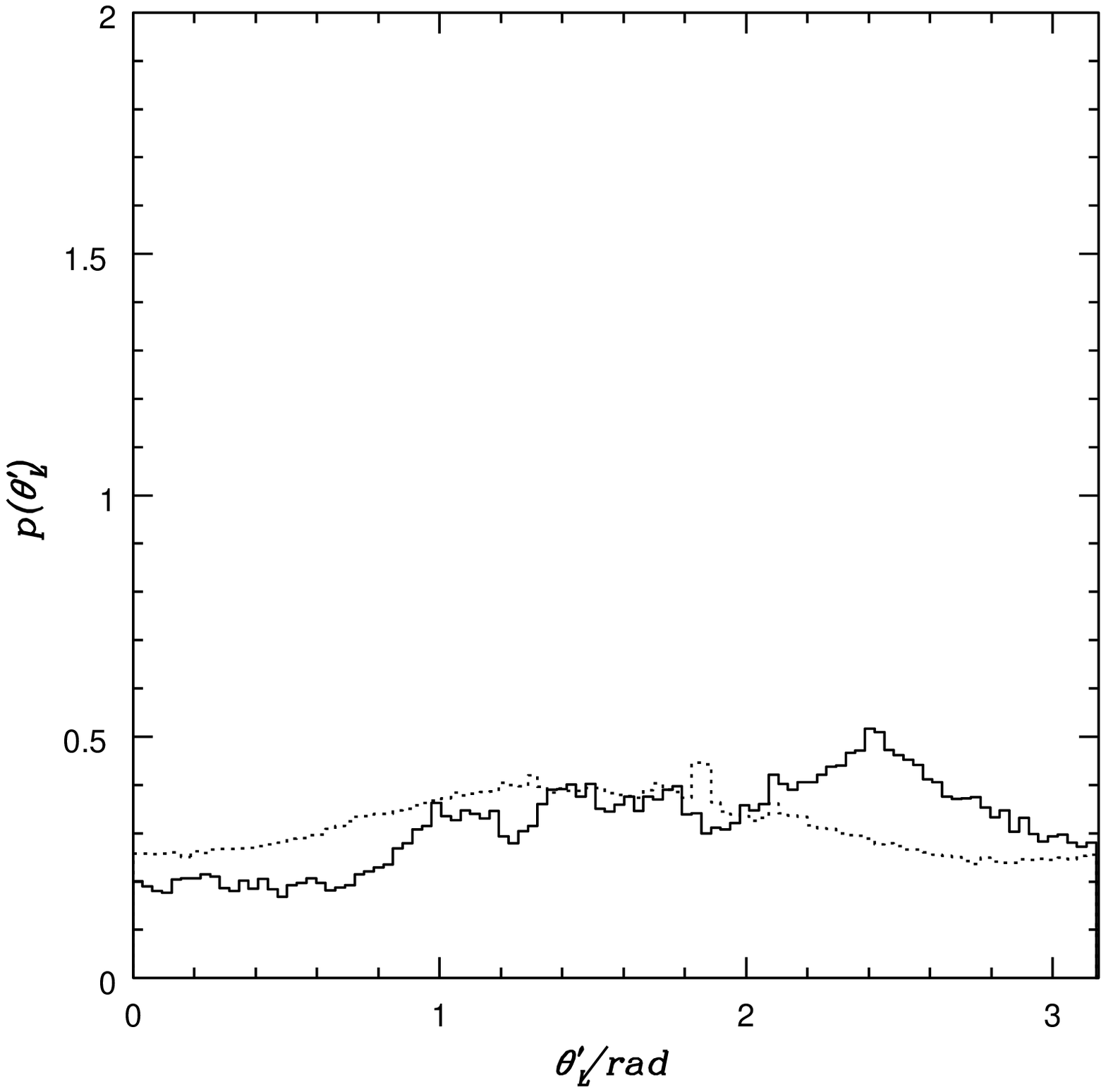}}
}
\mbox{
\scalebox{0.3}{\includegraphics{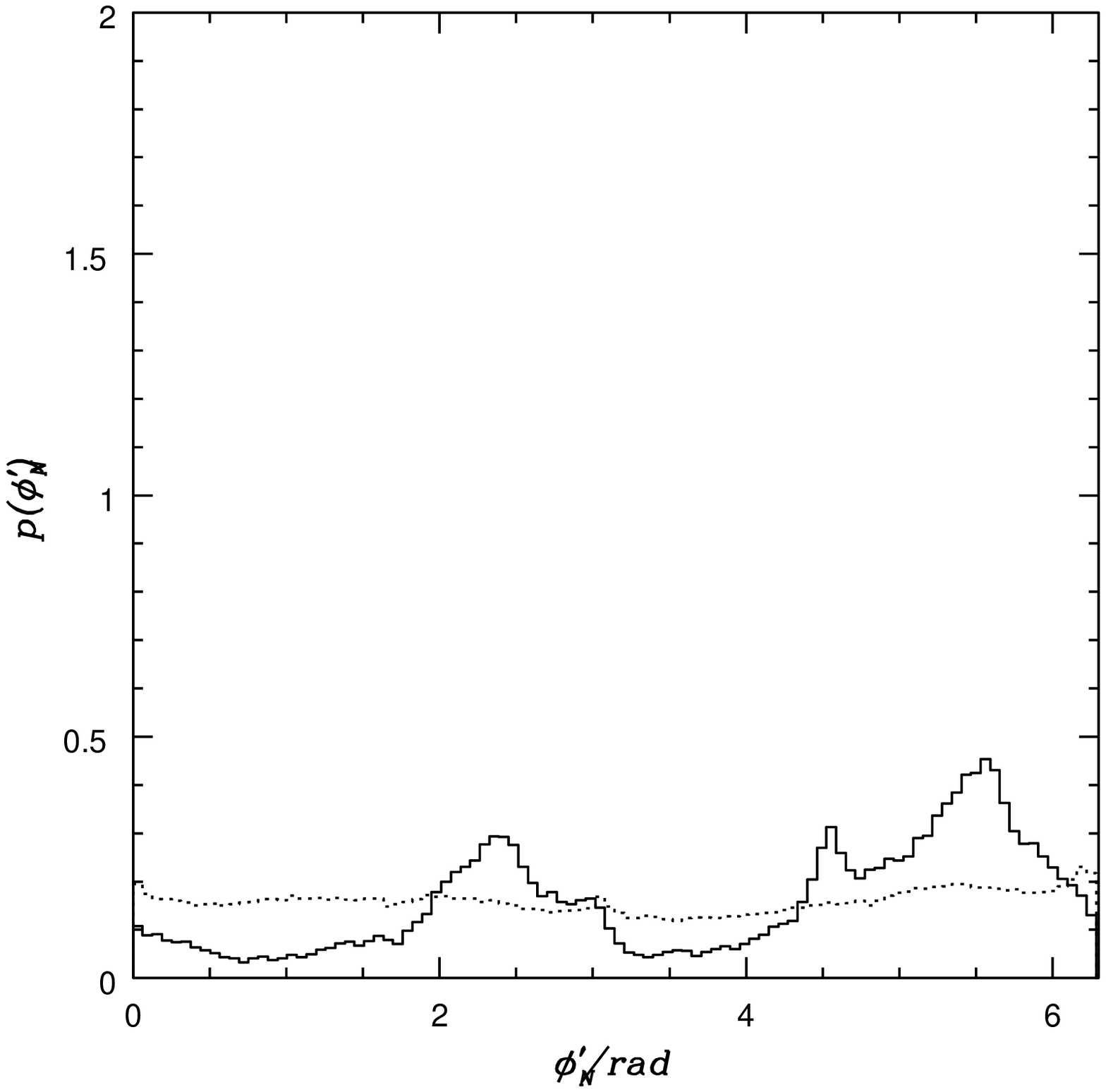}}
\scalebox{0.3}{\includegraphics{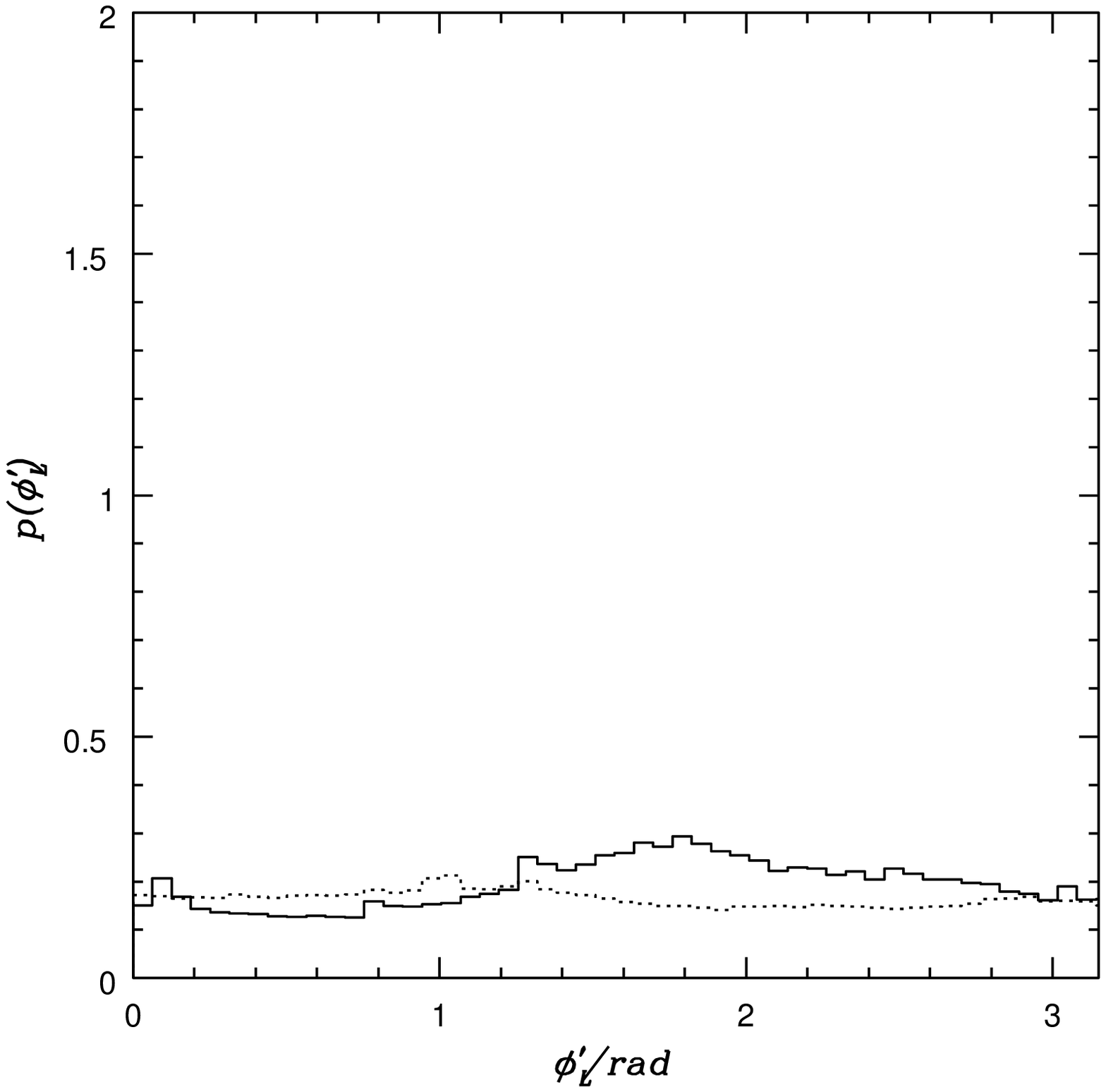}}
}
\caption{The posterior probability density functions for the four angular parameters describing the position $\hat N^{\prime}$ and orientation $\hat L^{\prime}$ of the DWD. The actual values of the DWD parameters used for the signal injection are reported in Table~\ref{t:param}. The two lines in the plots correspond to the MCMC results for the model $m_{B + W}$ (solid line) and the model $m_W$ (dotted line). Notice that here we do not show the relevant marginalised PDF's obtained from the multi-variate Gaussian distribution that one expects theoretically in the limit of signal-to-noise ratio $\rightarrow \infty$ because those parameters are essentially degenerate and we computed the Fisher information matrix only for the 3-dimensional case in which the parameters are $A_0,\ f_0,\ \phi_0^{\prime}$.}
\label{Fig_WDangles}
\end{center}
\end{figure}

Given the data set that we have just described, we compute/search for the (marginalised) posterior PDF's for each of the model parameters and the noise. We consider three different models assuming in each case that noise is also present (while the data set does not change): (i) a model consisting of a MBHB and a DWD ($m_{B+W}$), that corresponds to a truthful representation of the actual signals in the data set {\em c.f.}~(\ref{e:d}), hence we search over the vector of 15+1 unknown parameters $\vec{\Theta}_{B+W+n} = \{\vec{\theta}_B, \vec{\theta}_W, \sigma\}$; (ii) a model that contains only a MBHB ($m_B$) implying a search over the vector of 8+1 unknown parameters $\vec{\Theta}_{B+n} = \{\vec{\theta}_B, \sigma\}$; and (iii) a model that contains only a DWD ($m_W$) implying a search over the 7+1 parameter vector $\vec{\Theta}_{W+n} = \{\vec{\theta}_W, \sigma\}$. Using model $m_{B + W}$ we can explore whether we can study radiation from a DWD even when the stronger signal produced by the MBHB is present. Results obtained using either $m_B$ or $m_W$ provide us with useful complementary information about the behaviour of this choice of approach to the analysis. In order to quantitatively characterise the results, we also compute the variance-covariance matrix which allows us to evaluate the joint posterior PDF of the parameters in the limit $\rho \rightarrow \infty$; this yields a multi-variate Gaussian distribution centred on the true parameter values and should be compared to our numerical evaluation of Eq.~\ref{Bayestheorem} for the problem at hand. The results of our analysis are summarised in Figures \ref{Fig_BH}, \ref{Fig_WD+noise}, \ref{Fig_BHangles} and \ref{Fig_WDangles}. 

We consider first the results obtained using the (correct) model for the data, that is $m_{B+W}$. The first key point to note is that the unknown parameters $\vec{\Theta}_{B+W+n}$ (which includes the noise) are correctly recovered by our analysis scheme in which we simultaneously search for both the strong and weak signal. This result -- which is however limited just to one choice of source parameters -- is an indication that there might not be the need {\em a priori} to remove very loud signals to study weaker ones. We also note that the marginalised posterior PDF's of each parameter are broadly consistent with what is predicted theoretically (which assumes $\rho \rightarrow \infty$): the spread of the distributions is comparable to the value of the mean square errors as evaluated using the variance-covariance matrix, although the PDF's are not exactly Gaussian. This is expected in a real analysis process. As far as the individual PDF's are concerned, it is interesting to notice that although the observation time is just $\approx 12$ days, one can extract some information about the geometrical parameters that describe the MBHB. In general one would not expect this to be the case, because the modulations induced by the LISA motion are still small. The result can explained by the fact that the signal-to-noise ratio is large and the noise is constant (instead of increasing toward lower frequencies). On the other hand, no information can be gained about the source position and orientation of the orbital plane for the DWD (the PDF's for the parameters $\theta_N^{\prime},\ \phi_N^{\prime},\ \theta_L^{\prime},\ \phi_L^{\prime}$ are essentially flat over the entire relevant parameter space, and do not show any significant peak): this is easily explained by the fact that $f_0 = 5\times 10^{-4}$ Hz and over the short observation time the modulations (dominated by the LISA change of orientation) are too small, given the signal-to-noise ratio, to allow the disentangling of the geometrical parameters. This can also be easily verified by calculating the variance-covariance matrix, which is essentially degenerate for the 7 parameter case of the DWD considered in the simulation. In fact, the theoretical results that we show in the figures are obtained computing the variance-covariance matrix only for the 3-dimensional case in which the parameters are $A_0^{\prime},\ f_0,\ \phi_0^{\prime}$.

It is also instructive to consider the case in which we compute the posterior PDF's assuming the wrong models, $m_{B}$ and $m_W$. In each case, the source included in the model is still identified. However the PDF's are significantly biased away from the value corresponding to the parameters characterising the injected signals and the structure of the PDF's themselves are affected. The most dramatic change is however in the PDF of the variance of the noise, Fig.~\ref{Fig_WD+noise}, which is peaked significantly above the actual value $\sigma = 1$. This is a result of the algorithm being forced to assign the extra power from the un-searched for source to the noise. Such an effect is naturally stronger for $m_W$ where the entire contribution of the loud MBHB is left to be accounted for by the noise. 

\section{Conclusions}
\label{s:conclusions}

We have explored the application of Bayesian inference to the analysis of LISA data in the presence of gravitational waves emitted by sources of different classes. More specifically we have investigated whether the presence of strong radiation from massive black hole binary systems indeed prevents, if not previously removed from the data set, the study of weaker sources (such as white dwarf binaries). Our approach is to ``simultaneously fit'' the source parameters, by constructing the (marginalised) posterior probability density functions of the unknown signal model parameters, via an automatic Metropolis-Hastings (RJ) Markov Chain Monte Carlo sampler~\cite{Stroeer2006b}. The results of this preliminary study show that this approach is promising and could be suitable for LISA data analysis; it allows us to fully exploit the data without the need for adopting a recursive approach in which stronger signals are identified and progressively removed from the data stream (which is a very delicate process) in order to study weaker ones.

The results presented in this paper should be regarded only as a promising initial step of a more detailed exploration of this important analysis problem within the Bayesian framework. In fact, our analysis suffers from a number of limitations -- such as short integration time, simple waveforms, one Michelson observable, white noise -- that are derived from the need to keep the numerical implementation at a sufficiently simple level and to minimise the computational burden. We are currently extending the exploration of this analysis scheme to more realistic situations and, in order to do so, improving the efficiency of our code. Nonetheless, the technique that we have presented here is general and can be applied to a wide range of analysis challenges provided by LISA. 

\section*{Acknowledgements}

We would like to thank C.~Messenger, J.~Veitch and G.~Woan for several enlightening discussions about Bayesian inference, Markov Chain Monte Carlo techniques and relevant numerical implementation issues.

\section*{References}

\providecommand{\newblock}{}


\begin{thebibliography}{1}
\expandafter\ifx\csname url\endcsname\relax
  \def\url#1{{\tt #1}}\fi
\expandafter\ifx\csname urlprefix\endcsname\relax\def\urlprefix{URL }\fi
\providecommand{\eprint}[2][]{\url{#2}}

\bibitem{Bender98}
Bender, B.~L.,~et al., 
{\it LISA Pre-Phase A Report; Second Edition} 1998, MPQ 233.

\bibitem{Cutler2002a}
Cutler C and Thorne K~S 2002 arXiv:gr-qc/0204090

\bibitem{Ses2005a}
{Sesana} A, {Haardt} F, {Madau} P and {Volonteri} M 2005 {\em \apj\/} {\bf 623}
  23
\bibitem{Blanchet2002}
Blanchet L 2002  Living Rev.\ Rel.\  {\bf 5} 3

\bibitem{HBW90}
Hils, D., Bender,P.~L., \& Webbink, R.~F., ApJ {\bf 360}, 75 (1990).

\bibitem{NYPZ04}
Nelemans, G., Yungelson, L.~R., \& Portegies Zwart, S.~F., MNRAS 349,
181 (2004)

\bibitem{MNS04}
Marsh, T.~R.,  Nelemans, G., \&  Steeghs, D., MNRAS 350, 113-128 (2004).

\bibitem{BC2004}
Barack, L and Cutler, C (2004)  Phys.\ Rev.\ D {\bf 69} 082005

\bibitem{Cutler1998a}
{Cutler} C (1998) Phys.\ Rev.\ D {\bf 57} 7089--7102

\bibitem{Hughes2002}
Hughes S~A (2002) Mon.\ Not.\ Roy.\ Astron.\ Soc.\  {\bf 331} 805

\bibitem{Vecchio2004}
Vecchio A (2004) Phys.\ Rev.\ D {\bf 70} 042001

\bibitem{CL03}
{Cornish} N~J and Larson S~L 2003 Phys.\ Rev.\ D {\bf 67} 103001

\bibitem{Um2005a}
{Umst{\"a}tter} R, {Christensen} N, {Hendry} M, {Meyer} R, {Simha} V, {Veitch}
  J, {Vigeland} S and {Woan} G 2005 {\em Classical and Quantum Gravity\/} {\bf
  22} 901--+

\bibitem{Um2005b}
{Umst{\"a}tter}, R. and {Christensen}, N. and {Hendry}, M. and 
	{Meyer}, R. and {Simha}, V. and {Veitch}, J. and {Vigeland}, S. and 
	{Woan}, G. 2005 {\em \prd\/} {\bf 72} 022001

\bibitem{Corn2005a}
{Cornish} N~J and {Crowder} J 2005 {\em \prd\/} {\bf 72}(4) 043005

\bibitem{Stroeer2006b}
{Stroeer} A and {Vecchio} A, in preparation 

\bibitem{Green2003a}
{Green} P, {Hjort} N and {Richardson} S 2003 {Highly Structured Stochastic Systems}, Oxford University Press
  \urlprefix\url{citeseer.ist.psu.edu/594536.html}

\bibitem{Vecc2004a}
{Vecchio} A and {Wickham} E~D 2004 {\em \prd\/} {\bf 70}(8) 082002--+

\bibitem{TD05}
Dhurandhar, S.~V., \& Tinto, M, Liv. Rev. Rel. 8, 4 (2005)



\end{thebibliography}
\end{document}